%% file: STARMuonID.tex
\documentclass[a4paper,10pt,twocolumn,preprint,3p]{elsarticle}
\graphicspath{{plots/}{plotsINT/}}
\usepackage{graphicx, epsfig, subfigure}
\usepackage{subfigure}
\usepackage{mathrsfs}
\usepackage{longtable}
\usepackage{rotating}
\usepackage{graphicx}

\usepackage{lineno}

\usepackage{amssymb}
\journal{NIMA}

\begin{document}

\begin{frontmatter}



\date{\today}
    
\title{ {\bf Muon Identification with Muon Telescope Detector at the STAR Experiment } }


\author[ncku]{T.C. Huang}
\author[bnl]{R. Ma}
\author[uic]{B. Huang}
\author[thu]{X. Huang}
\author[bnl]{L. Ruan}
\author[bnl]{T. Todoroki}
\author[bnl]{Z. Xu}
\author[ustc]{C. Yang}
\author[ustc]{S. Yang}
\author[ustc]{Q. Yang}
\author[ncku]{Y. Yang \corref{cor1}}
\ead{yiyang@ncku.edu.tw}
\author[ustc]{W. Zha}

\address[ncku]{National Cheng Kung University, Tainan 70101, Taiwan}
\address[bnl]{Brookhaven National Laboratory, Upton, New York 11973, USA}
\address[thu]{Tsinghua University, Beijing 100084, China}
\address[uic]{University of Illinois at Chicago, Chicago, Illinois 60607, USA}
\address[ustc]{University of Science and Technology of China, Hefei 230026, China}

\cortext[cor1]{Corresponding author.}

\begin{abstract}
    The Muon Telescope Detector (MTD) is a newly installed detector in the STAR experiment. 
    It provides an excellent opportunity to study heavy quarkonium physics using the dimuon channel in heavy ion collisions. 
    In this paper, we report the muon identification performance for the MTD using proton-proton collisions at $\sqrt{s}$ = 500 GeV with various methods. 
    The result using the Likelihood Ratio method shows that the muon identification efficiency can reach up to $\sim$90\% for muons with transverse momenta greater than 3 GeV/c and the significance of the $J/\psi$ signal is improved by a factor of 2 compared to using the basic selection. 

\end{abstract}

\begin{keyword}
STAR  \sep MTD \sep muon identification \sep muon \sep dimuon \sep quarkonium



\end{keyword}

\end{frontmatter}



%\begingroup

%\linenumbers

%\let\clearpage\relax
\input{intro}
\input{detector}

\input{selections}

\input{muid}

\input{results}

\input{conclusion}

\input{acknowledgments}




%\bibliographystyle{elsarticle-num}
\section*{\refname}
\bibliographystyle{elsarticle-num}
\bibliography{STARMuonID}

%\endgroup

%\clearpage

%\begingroup
%\let\clearpage\relax

%%%%%%%% Appendix
%\onecolumn
%\appendix
%\noindent
%{\Large \bf APPENDICES}
%\include{app_embed}
%\include{app_n1}
%\include{app_llr}
%\include{app_datavsmc}

%%\endgroup


%\begin{thebibliography}{00}


%\bibitem{}

%\end{thebibliography}
\end{document}

%% file: intro.tex
\section{Introduction}
\label{intro}

The Solenoidal Tracker At RHIC (STAR)~\cite{star_det} is one of the two large high energy nuclear physics experiments at the Relativistic Heavy Ion Collider (RHIC) at Brookhaven National Laboratory. 
%After 15 years of operation, the STAR experiment provided us many important physics improving our understanding the properties of Quantum Chromodynamics (QCD). 
After 15 years of operation, the STAR experiment has provided many important results, which have helped to improve our understanding of Quantum Chromodynamics (QCD). 
In particular, evidence of the existence of the Quark Gluon Plasma (QGP) opened a new window to get a deeper insight into QCD~\cite{dis_qgp}. 
%Heavy quark-antiquark states are perfect probes to study the properties of the QCD matter, for example the colour screening effect, and dimuon channel can provide us much cleaner signals. 
Heavy quarkonia states are ideal probes to study the properties of QCD matter.
For instance, quarkonium suppression in the medium due to the color-screening of surrounding partons can provide information about the partonic nature of the QGP and its temperature~\cite{jpsi_supp, pot_mod}. 
%To reconstruct the quarkonium states via the leptonic decay, the dimuon channel suffers from less combinatoric background and Bremsstrahlung radiation compared to the dielectron channel.
%Quarkonia are identified by their reconstructed mass peak in the dilepton decay channel. 
Quarkonia are identified by reconstructing their invariant mass ($M_{inv.}$) in the dilepton decay channel. 
Muons can be reconstructed more precisely due to their reduced bremsstrahlung radiation in material compared to electrons. 
%Therefore, identifying muon candidates plays an important role in heavy quarkonium physics.
%There are many studies on this subject from different experiments; more details can be found in Refs.~\cite{muid_atlas, muid_cms, muid_lhcb}.
%many experiments have dedicated study on this subject~\cite{muid_atlas, muid_cms, muid_lhcb}. 
%The Muon Telescope Detector (MTD)~\cite{mtd, mtd_2} was started to install and to operate in the STAR detector since 2012, and the full installation was completed in 2014. 
A new subdetector in STAR, the Muon Telescope Detector (MTD), dedicated to measuring muons was proposed in 2009 and was installed from 2012 to 2014~\cite{mtd, mtd_2}.
%The installation of Muon Telescope Detector (MTD)~\cite{mtd, mtd_2} was started from 2012 to 2014, and it has been operating since 2012.
In this paper, we present the muon identification performance of this new detector. 
There have been many studies on muon identification from different experiments, and more details can be found in Refs.~\cite{muid_atlas, muid_cms, muid_lhcb}.

This paper is arranged as follows. 
In Section 2, a brief description of the STAR detector is presented. 
The data sets and event selection are described in Section 3.
In Section 4, three methods used to identify muon candidates are described. 
We present in section 5 the efficiency as well as the resulting signal significance for $J/\psi$ from these three methods. 
Finally, a summary is given in Section 6. 

%% file: detector.tex
\section{The STAR detector}
\label{star}
The STAR detector is a general purpose particle detector optimized for high energy nuclear physics. 
%The subsystems include the Heavy Flavor Tracker (HFT) which is the inner most semiconductor-based tracking system for reconstructing the secondary vertex from B-hadron decay~\cite{hft}; 
The main subsystems relevant to this analysis include the Time Projection Chamber (TPC), the Magnet System and the MTD. 
The TPC is the primary tracking detector for charged particles and provides particle identification via measurements of the energy loss ($dE/dx$)~\cite{tpc}. 
It covers full azimuthal angles ($0 < \phi < 2 \pi$) and a large pseudorapidity range ($|\eta| < 1$). 
%the Time of Flight detector (TOF) was installed and commissioned in 2010 and it improved the particle identification to higher energy ($p \sim 3$ GeV/c)~\cite{tof, tof_2}; 
%The Barrel Electromagnetic Calorimeter (BEMC) is installed outside the TOF detector and it provides triggers for particles with electromagnetic interaction and precision measurements in energy~\cite{bemc}; 
%The Magnet System generates a 0.5 Tesla solenoidal field to bend the charged particles.
The transverse momenta ($p_T$) and charge (q) of charged particles are measured by the curvature of their trajectories in the 0.5 Tesla solenoidal field generated by the Magnet System.
%The Magnet System generates a 0.5 Tesla solenoidal field; and the transverse momenta ($p_T$) of charged particles are measured based on their trajectories in this field. %transverse momentum ($p_T$) measurement is based on the bending of charged particle trajectories in the magnet field.
There are 30 bars, known as ``backlegs'', outside the coil to provide the return flux path for the magnetic field~\cite{magnet}. 
They are 61 cm thick at a radius of 363 cm corresponding to about 5 absorption lengths.
These backlegs play an essential role in enhancing the muon purity by absorbing the background hadrons from collisions.
%and Muon Telescope Detector (MTD) is based on the Multi-gap Resistive Plate Chambers (MRPC) technology to record the hits from charged-particles in short time~\cite{mtd}. 
%This magnet is cylindrical and consists of 30 flux-return bars, four end rings, and two pole tips. The return flux path for the field is provided by the outer magnet steel [27]. The 6.85 m long flux-return bars, also called “backlegs,” are trapezoidal in cross-section and 57 cm thick at a radius of 363 cm. These return bars and the BEMC serve as a hadron absorber allowing only muons to reach the MTD in elementary and heavy- ion collisions.
The MTD is a fast detector based on the Multi-gap Resistive Plate Chamber technology to record signals, also referred to as signals (``hits'') generated by charged particles traversing it. 
%It also plays an important role in the High Level Trigger system (HLT) for triggering muon candidates. 
%The muon triggers in STAR HLT system include the single-muon, dimuon and electron-muon trigger for various physics interests. 
It provides single-muon and dimuon triggers based on the number of hits within a predefined online timing window. 
%The acceptance covers about 45\% in azimuth ($\phi$) within $|\eta| <$ 0.5. 
%The MTD modules are installed at a radius of 403 cm and provide one space point measurement for each track. 
%and it covers azimuthal angle from 0 to 2$\pi$ and pseudorapidity up to 0.5~\cite{mtd}.   
The MTD modules are installed at a radius of about 403 cm, and cover about 45\% in azimuth within $|\eta| <$ 0.5~\cite{mtd}.
%The MTD detector was installed 10\%, 63\% and 100\% of the total coverage in 2012, 2013, and 2014. 
Installation of the full MTD was 10\%, 63\%, and 100\% completed for the 2012, 2013, and 2014 run years respectively.
%Out of the full MTD system, 10\%, 63\% and 100\% was installed in 2012, 2013, and 2014, respectively.
%The timing and spatial resolution of the MTD obtained from cosmic ray data are $\sim$100 ps and $\sim$1-2 cm in $\phi$ and z direction, respectively~\cite{mtd_3}.  
%As shown in cosmic ray data, the timing and spatial resolutions of the MTD are $\sim$100 ps and $\sim$1-2 cm in both $\phi$ and z directions, respectively~\cite{mtd_3}.
As shown in cosmic ray data, the timing resolution of the MTD is $\sim$100 ps and the spatial resolutions are $\sim$1-2 cm in both $r\phi$ and $z$ directions~\cite{mtd_3}.

%% file: selections.tex
\section{Dataset and event selection}
\subsection{Data and Monte Carlo}
Data for this study were collected by the STAR detector during the RHIC proton-proton run at a center of mass energy of 500 GeV in 2013. 
Events in the data sample were selected using the MTD dimuon trigger which requires at least two MTD hits in coincidence with the bunch crossing. 
%and the corresponding integrated luminosity is 28.3 $pb^{-1}$. 
The data set represents an integrated luminosity of 28.3 pb$^{-1}$.
%After the basic track selection described in the next section, more than 1500 $J/\psi \to\mu^+\mu^-$ candidates are used in this study. 

%In order to have pure and larger statistics muon sample, the Monte Carlo of the $J/\psi \to \mu^+ \mu^- $ signal was used in this study.  
%The sample was generated by a single-particle generator with flat distributions in $J/\psi$ $p_T$, $\phi$ and $\eta$. These simulated signals were then passed through the full GEANT3 simulation of the STAR detector, and ``embedded'' into real events followed by the standard reconstruction procedure as used for real data.
The detector response to the $J/\psi \to \mu^+ \mu^- $ signal was studied using a Monte Carlo (MC) simulation. 
The MC sample was generated by a single-particle generator with flat distribution in $p_T$, $\phi$ and $\eta$ for $J/\psi$. 
These simulated signals were then passed through the full GEANT3~\cite{geant3} simulation of the STAR detector, and ``embedded'' into real events, followed by the standard reconstruction procedure as used for real data.
The kinematic distributions of the embedded $J/\psi$ and $\mu^{\pm}$ were weighted by the $p_T$ spectrum of $J/\psi$ in $pp$ collisions at 500 GeV determined via interpolation through a global fit of world-wide differential $J/\psi$ cross section measurements~\cite{jpsi_pt}.
%In order to obtain the correct shape of reconstructed $J/\psi$ mass distribution in MC, the reconstructed muon $p_T$ was smeared by a Gaussian function as described in Eq.\ref{eq:pt_smear}. 
The reconstructed muon $p_T$ in MC was also slightly smeared by a Gaussian function, with mean = $1.004\times p_T$ and width = $0.022\times p_T$, to match the reconstructed $J/\psi$ mass distribution in data.%to obtain a more realistic reconstructed $J/\psi$ mass distribution. 
%\begin{equation}
%    \label{eq:pt_smear}
    %f(p_T) = \frac{1}{\sigma\times\sqrt{2\pi}}e^{ -\frac{ (p_T - \mu)^2}{2 \sigma^2}}
%    f(p_T) = Gaus( mean, width).
    %p_T^{tune} = \frac{1}{(p_T^{reco.}\times {\rm smear}) \sqrt{2\pi}} e^{-\frac{ (p_T^{reco.} - (p_T^{reco.}\times(1+{\rm shift})))^2}{2(p_T^{reco.}\times {\rm smear})^2}}.
%\end{equation}
%where mean and width are $p_T^{reco.}\times(1+{\rm shift})$ and $p_T^{reco.}\times {\rm smear}$, respectively. 
%The correct kinematics of $J/\psi$ and $\mu^{\pm}$ were obtained by an event-by-event re-weighting procedure on each truth event. 
%The weight is determined by the ratio of the $p_T$ of $J/\psi$ from MC in the truth level and the global fit of $J/\psi$ cross section from world-wide measurements~\cite{jpsi_pt}.
%The weight is determined by the ratio of the $p_T$ distribution of $J/\psi$ from MC to a true $p_T$ distribution of $J/\psi$ derived from  world-wide measurements by a global fit~\cite{jpsi_pt}.
%To get a more realistic simulation sample, the signal events were "embedded" into real data from STAR to describe the background contribution. 

%The Monte Carlo sample used in this study is $ pp \to J/\psi X \to \mu^+ \mu^- X$ generated by {\sc Pythia6}~\cite{pythia6} with full simulation of the STAR detector.
%To get a more realistic simulation sample, the signal events were "embedded" into real data from STAR to describe the background contribution. 

\subsection{Track selection}
Tracks selected for the muon identification study have to meet the following requirements: %the tracks were from the primary vertex in the event; 
$p_T$ is greater than 1 GeV/c; 
%the psuedorapidity of the track must be within the MTD coverage, $|\eta| < 0.5$; 
the distance of closest approach to the collision vertex should be less than 3 cm to suppress secondary decays; 
number of TPC clusters used in reconstruction should be greater than 15 (the maximum possible is 45) to have good momentum resolution; 
number of TPC clusters used for the $dE/dx$ measurement is greater than 10 to ensure good $dE/dx$ resolution;
the ratio of the number of used TPC clusters over the number of possible clusters is not less than 0.52 in order to reject split tracks. 
%the track should match to a valid TOF hit with the projected position within TOF's sensitive readout volume.
Tracks are also required to project to MTD hits that fire the triggers. 
%Tracks are also required to match the MTD hits that fire the triggers. 
%The candidates selection are summarized in Table~\ref{table:dimuon_selection}.
%\begin{table}[htbp]
%  \begin{center}
%  \begin{tabular}{ll}
%  %\hline \hline
%  %    Variable            &   Requirement  \\
%  \hline\hline
%  %\multicolumn{2}{c} { Basic requirements } \\
%  %\hline
%  Primary tracks   &       \\
%  $p_{\rm T}$    &      $>$ 1.0 GeV/c \\
%  $|\eta|$  & $<$ 0.8 \\
%  DCA   & $<$ 3 cm     \\
%  nHits-Fits &    $\geq$ 15     \\
%  nHits-dEdx &   $\geq$ 10     \\
%  rHits-all &    $\geq$ 0.52    \\
%  MTD hits &      \\
%  \hline \hline
%   \end{tabular}
%  \end{center}
%  \caption{A summary of the basic selection criteria for muon candidates.}
%\label{table:dimuon_selection}
%\end{table}

\begin{figure}[!htbp]
  \begin{center}
      \includegraphics[width=0.48\textwidth]{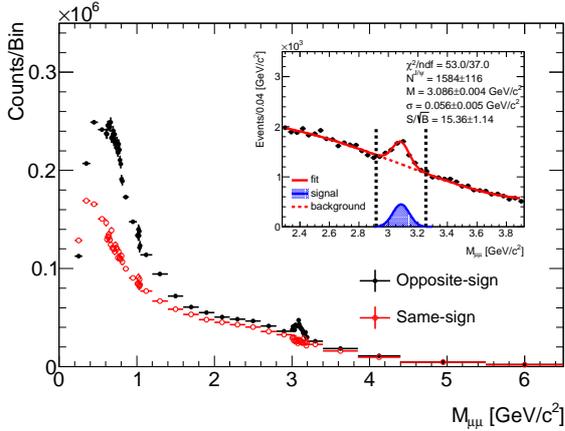}
      \end{center}
  \caption{  The dimuon mass spectrum with basic selection described in Section 3.2 applied to both muons. The black solid and red open circles are for the muon pairs with opposite and same signs in charge, respectively. The inset shows the dimuon mass fits for $J/\psi$ peak. The solid red line is a combined fit to the signal and background with a single Gaussian plus a fourth-order polynomial, and the two vertical black dashed lines indicate the mass window [2.92 - 3.25 GeV/c$^2$] used to select the $J/\psi$ candidates.
  \label{fig:dimuon_mass}}
\end{figure}

%Figure~\ref{fig:dimuon_mass} shows the invariant mass spectrum of opposite-sign dimuon pairs with the selection criteria described above applied to both muons and it is clear to see the $J/\psi$ signal around 3.1 GeV/c$^2$. 
Figure~\ref{fig:dimuon_mass} shows the invariant mass spectrum of opposite-sign dimuon pairs with the selection criteria described above applied to both candidate daughters. The $J/\psi$ signal is clear around 3.1 GeV/c$^2$. 
More than 1500 $J/\psi$ candidates are present in the data sample used here. 
Lighter mesons, like $\omega$, $\phi$ and $\eta$ particles particles are obscured by large backgrounds at low $M_{inv.}$.
%On the other hand, the light mesons, like $\omega$, $\phi$ and $\eta$ particles, are not visible due to the huge background. 

%% file: muid.tex
\section{Muon identification}
\label{muid}
\subsection{Methods}
\begin{figure}[htbp]
  \begin{center}
      \includegraphics[width=0.5\textwidth]{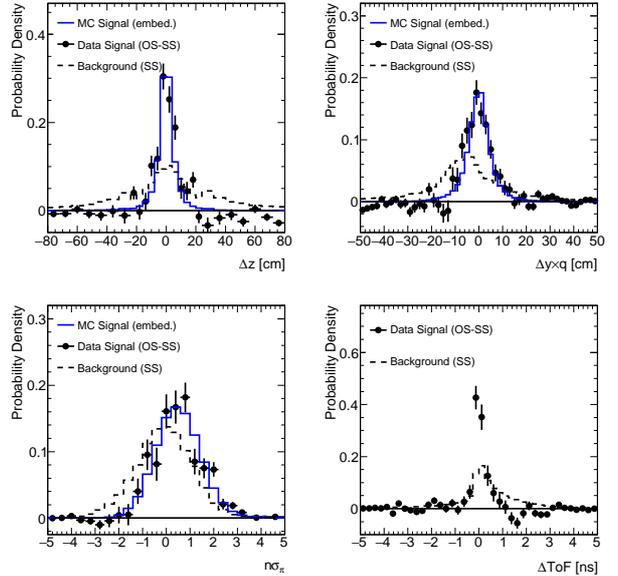}
      \end{center}
      \caption{ The probability density function of $\Delta$z, $\Delta y \times q$, $\Delta$ToF, and $n\sigma_{\pi}$ variables. The black points are the signal from data (OS-SS), the solid line histograms are the signal from MC, and the dashed line histograms are the background from data (SS). 
  \label{fig:pdf}}
\end{figure}

%In order to obtain purer muon sample from MTD, $J/\psi\to\mu^+\mu^-$
%There are four variables, $\Delta$ToF, $\Delta y\times q$, $\Delta z$ and $\Delta n\sigma_{\pi}$, to distinguish the muon candidates from background, mainly hadrons (pions and kions) in STAR.  
To distinguish muon candidates from the hadron background, there are four variables, $\Delta$ToF, $\Delta y\times q$, $\Delta z$ and $n\sigma_{\pi}$ used in this study.  
$\Delta$ToF is the difference between the calculated time-of-flight value from track extrapolation with a muon particle hypothesis and the measured one from the MTD detector. 
$\Delta z$ and $\Delta y$ are the residuals between the MTD hit position and extrapolated track position on the MTD, where $z$ is along the beam pipe and $y$ is perpendicular to z along the surface of each MTD module (approximately $r\phi$). 
%Since $\Delta y$ distribution depends on the charge, 
%To eliminate the charge dependence in the $\Delta y$ variable, therefore the quantity $\Delta y\times q$ were used. 
$\Delta y$ is multiplied by charge ($\Delta y\times q$) to eliminate the charge dependence. 
%$n\sigma_{\pi}$ is the normalized $dE/dx$ quantity to the theoretical value of pions, defined as 
$n\sigma_{\pi}$ is the difference between the measured $dE/dx$ and the theoretical value assuming the track is a pion (for simplicity with pre-existing codes), normalized to the $dE/dx$ resolution of the TPC: 
\begin{equation}
    n\sigma_{\pi} = \frac{ (\log \frac{dE}{dx})_{{\rm measured}} - (\log \frac{dE}{dx})_{\pi, {\rm theory}} }{\sigma( \log \frac{dE}{dx})_{{\rm measured}} }.
\end{equation}
%where $\sigma( \log \frac{dE}{dx})_{{\rm measured}}$ is the dE/dx resolution of the TPC.
%In order to obtain the normalized distribution, as known as the probability density function (PDF), of each variable for pure muons, the muon events under $J/\psi \to \mu^+\mu^-$ peak were used as shown in the insert of Fig.~\ref{fig:dimuon_mass}, where the two vertical black dashed lines indicate the mass window [2.92 - 3.25 GeV] used to select the muon sample.
%To obtain the probability density function (PDF) for pure muons, muon events under the $J/\psi$ peak were used as shown in the insert of Fig.~\ref{fig:dimuon_mass}, where the two vertical black dashed lines indicate the mass window, i.e. [2.92, 3.25] GeV/c$^2$, used to select the muon sample.
%To compare the signal and background distributions for each variable, the probability density functions (PDFs) for pure muons are obtained from the muon events under the $J/\psi$ peak as shown in the insert of Fig.~\ref{fig:dimuon_mass}, where the two vertical black dashed lines indicate the mass window, i.e. [2.92, 3.25] GeV/c$^2$, used to select the muon sample.
%The signal PDFs are extracted using opposite-sign (OS) events subtracting out the same-sign (SS) events, while the PDFs for background are simply using the same-sign events in the same mass region.
The probability distribution functions (PDFs) of each variable for same sign dimuon pairs (SS) within the $M_{inv.}$ window [2.92, 3.25] GeV/c$^2$ (shown by vertical dashed black lines in the inset of Fig.~\ref{fig:dimuon_mass}) were used to characterize backgrounds. 
The PDFs of pure muons were then obtained by a subtraction of the background PDFs from those of opposite sign dimuon pairs (OS) within the same $M_{inv.}$ window. 
%However, the $\Delta$ToF distribution in MC was tuned to match the signal in data due to the lack of information in simulation.
Similar distributions are extracted from MC as well except for the $\Delta$ToF distribution because the timing signal of MTD is not modeled in the simulation.
%Due to the limited statistics in data for the signal PDFs which causes large fluctuations, the embedded MC sample is used to construct the signal PDFs.
Figure~\ref{fig:pdf} shows the comparisons for the PDF of each variable between signal (data and MC) and background. 
The signal distributions in data and MC are in reasonable agreement. 

Three methods, straight cut, N-1 iteration and Likelihood Ratio, are utilized in this paper to identify muon candidates. 
The performance of these three methods is quantified using a tag-and-probe procedure. 
In the low muon $p_T$ region ($p_T < 3.5$ GeV/c), the tagged muon is the one with higher $p_T$, while the probed muon is the one with lower $p_T$. %, and vice versa in the high $p_T$ region ($p_T < 3.5$ GeV/c).
However, in the high muon $p_T$ region ($p_T > 3.5$ GeV/c), in contrast, the tagged (probed) muon is the one with lower (higher) $p_T$ to increase statistics.
The muon identification cuts are applied on the probed muons, and then the efficiency and background rejection power can be obtained by comparing the yield and significance of the $J/\psi$ signal from before and after the selection. 
Several signal (single Gaussian or Crystal-Ball function) and background (third-order or fourth-order polynomial function) models are used to fit the dimuon mass spectra.
The efficiency is calculated by using the average fitted values for the number of $J/\psi$ from all combinations of signal and background models, and the fit model uncertainty is determined by using the maximum deviation of any fit result from the average. 
\begin{itemize}

\item {\bf Straight cut method} \\
%One way to reduce background is to simply apply straight cuts on these four variables independently. 
%The cut values for $|\Delta z|$, $|\Delta y \times q|$ and $n\sigma_{\pi}$ are determined through the widths of the signal distributions, and an asymmetric cut is used for $\Delta{\rm ToF}$.  
%The simplest way to reduce background is to straightly apply cuts on these four variables based on the widths of the signal distributions as shown in Fig.~\ref{fig:pdf}. 
The simplest way to reduce background is to directly apply cuts on these four variables. 
An about $\pm 2.5 \sigma$ window cut on $\Delta z$ and $\Delta y \times q$, and an asymmetric window cut\footnote{The mean value of $n\sigma_{\pi}$ for muons is shifted to the right by $\sim0.5\sigma$ compared to pions; therefore, a more strict cut on low $n\sigma$ is applied to reduce the pion contamination.}, $-1.5 \sigma$ to $+2.5 \sigma$, on $n\sigma_{\pi}$ are used to retain high efficiency while rejecting background, where $\sigma$ is the width of the signal distributions as shown in Fig.~\ref{fig:pdf}. 
%and an asymmetric cut is used for $\Delta{\rm ToF}$ from our experience.
An empirical asymmetric cut is used for $\Delta{\rm ToF}$ since the hadron background has a long tail to the right.
%The cut values for $|\Delta z|$, $|\Delta y \times q|$ and $n\sigma_{\pi}$ are chosen as $\pm 3\sigma$ of the widths, and an asymmetric cut is used for $\Delta{\rm ToF}$ from our experience. 
Specifically, the selection criteria are $-5 < \Delta{\rm ToF} < 0.2$ ns, $|\Delta z| < 20$ cm, $|\Delta y \times q| <20$ cm and $-1 < n\sigma_{\pi} < 3$.

\item {\bf N-1 iteration method } \\
    %A advanced way to select muon candidates is using $J/\psi \to \mu^+ \mu^-$ signal and iterating the cut values to obtain the best signal significance ($S/\sqrt{B}$) until the cuts are stable. 
    An advanced way to select muon candidates, called N-1 iteration, is to vary one variable to optimize the $J/\psi$ signal significance ($S/\sqrt{B}$) with the other N-1 variables fixed at each iteration step. 
%    iterate the cut values until the best $J/\psi$ signal significance ($S/\sqrt{B}$) is achieved and the cuts are stable. 
    The values of the cuts determined using this method are $-4.8 < \Delta{\rm ToF} < 0.7$ ns, $-15 < \Delta z < 19$ cm, $-9 < \Delta y \times q < 14$ cm and $-0.5 < n\sigma_{\pi} < 3.6$.

\item {\bf Likelihood Ratio method} \\
%An advanced way to reduce the background level and keep high purity simultaneously is using multivariate methods, and the Likelihood Ratio method is one of the most transparent multivariate methods. 
A more sophisticated way to reduce the background level and keep high purity simultaneously is using more powerful multivariate methods, such as the Likelihood Ratio method. 
The basic idea is to create a discriminative variable in the form of a likelihood ratio $R=(1-Y)/(1+Y)$, where $Y = \prod y_i$ and each $y_i=PDF^{\rm bkg}_i / PDF^{\rm sig}_i$ is a ratio between background and signal PDFs. %probability density functions (PDFs).
%In order to obtain the PDF of each variable for pure muons, the events under the $J/\psi \to \mu^+\mu^-$ peak were used as shown in the insert of Fig.~\ref{fig:dimuon_mass}, where the two vertical black dashed lines indicate the mass window used to select the muon sample.
%The signal PDFs are extracted using opposite-sign (OS) events subtracting out the same-sign (SS) events, while the PDFs for background are simply using the same-sign events in the same mass region.  
Due to the limited statistics in data for the signal PDFs which causes large fluctuations, the embedded MC sample is used to construct the signal PDFs.
%%However, the distributions have more fluctuation due to the limited statistics in data, so the PDFs from MC are used for signal.
%%Figure~\ref{fig:pdf_deltaz} shows an example of the comparison in the PDF of $\Delta z$ variable between signal (data and MC) and background.
%Figure~\ref{fig:pdf} shows the comparisons for the PDF of each variable between signal (data and MC) and background. 
%The signal distributions in data and MC are in good agreement. 
%\begin{figure}[htbp]
%  \begin{center}
%      \includegraphics[width=0.5\textwidth]{pdf}
%      \end{center}
%  \caption{ The probability density function of $\Delta$z, $\Delta y \times q$, $\Delta$ToF, and $n\sigma_{\pi}$ variables. The black points are the signal from data (OS-SS), the blue histograms are the signal from MC, and the black histograms are the background from data (SS). 
%  \label{fig:pdf}}
%\end{figure}
%We obtained the signal PDFs from signal Monte Carlo, and background PDFs from the same-sign data.
%Figure~\ref{fig:LHratio_muid} shows discriminating power and the signal efficiency versus background rejection rate for $J/\psi$ signal by using Likelihood Ratio selection. 
In this method, only three variables, $\Delta y\times q$, $\Delta z$ and $n\sigma_{\pi}$, are used to calculate the $R$ value for the probed muon. 
The cut values on $\Delta$ToF are fixed from the N-1 iteration method. % due to the incorrect correlations between $\Delta$ToF and other variables in simulation.  %mismatch of signal PDF between data and simulation.
Figure~\ref{fig:pdf_ratio} shows the PDF ratios ($PDF^{bkgd}/PDF^{sig}$) for, $\Delta y \times q$, $\Delta z$, and $n\sigma_{\pi}$, respectively. 
A bin-to-bin interpolation (solid red line) is used to obtain the ratios between points in the middle region while a linear fit (dashed red line) is used for the side regions where statistics are low.
Figure~\ref{fig:mu2_Likelihood_ratio} shows the discriminating power of the Likelihood Ratio method, and the cut value on $R$ variable, $R > -0.2$, is chosen to maximize the significance of the $J/\psi$ signal as shown in Fig.~\ref{fig:mu2_best_R_cut}.
\begin{figure}[htbp]
  \begin{center}
      \includegraphics[width=0.5\textwidth]{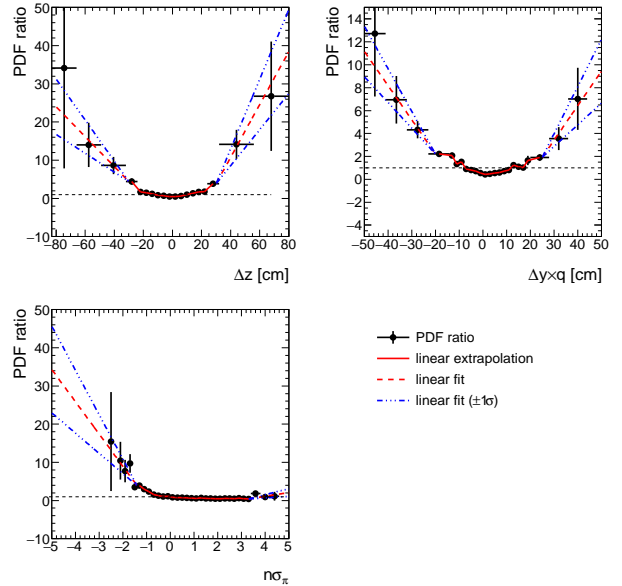}
      \end{center}
      \caption{ The PDF ratios of $\Delta$z, $\Delta y \times q$, $\Delta$ToF, and $n\sigma_{\pi}$ variables. The solid red line indicates the bin-to-bin interpolation and the dashed red line is a linear fit to parameterize the ratio.  
  \label{fig:pdf_ratio}}
\end{figure}

%\begin{figure}[!htbp]
%  \begin{center}
%      \includegraphics[width=0.4\textwidth]{mu2_Likelihood_ratio}
%  \end{center}
%  \caption{ Distributions of the Likelihood Ratio (R) for signal (blue) and background (red). The cut value on R variable is optimized by the signal significance.    
%  \label{fig:LHratio_muid}}
%\end{figure}

\begin{figure*}[htbp]
  \begin{center}
    \subfigure[]{
      \label{fig:mu2_Likelihood_ratio}
      \includegraphics[width=0.4\textwidth]{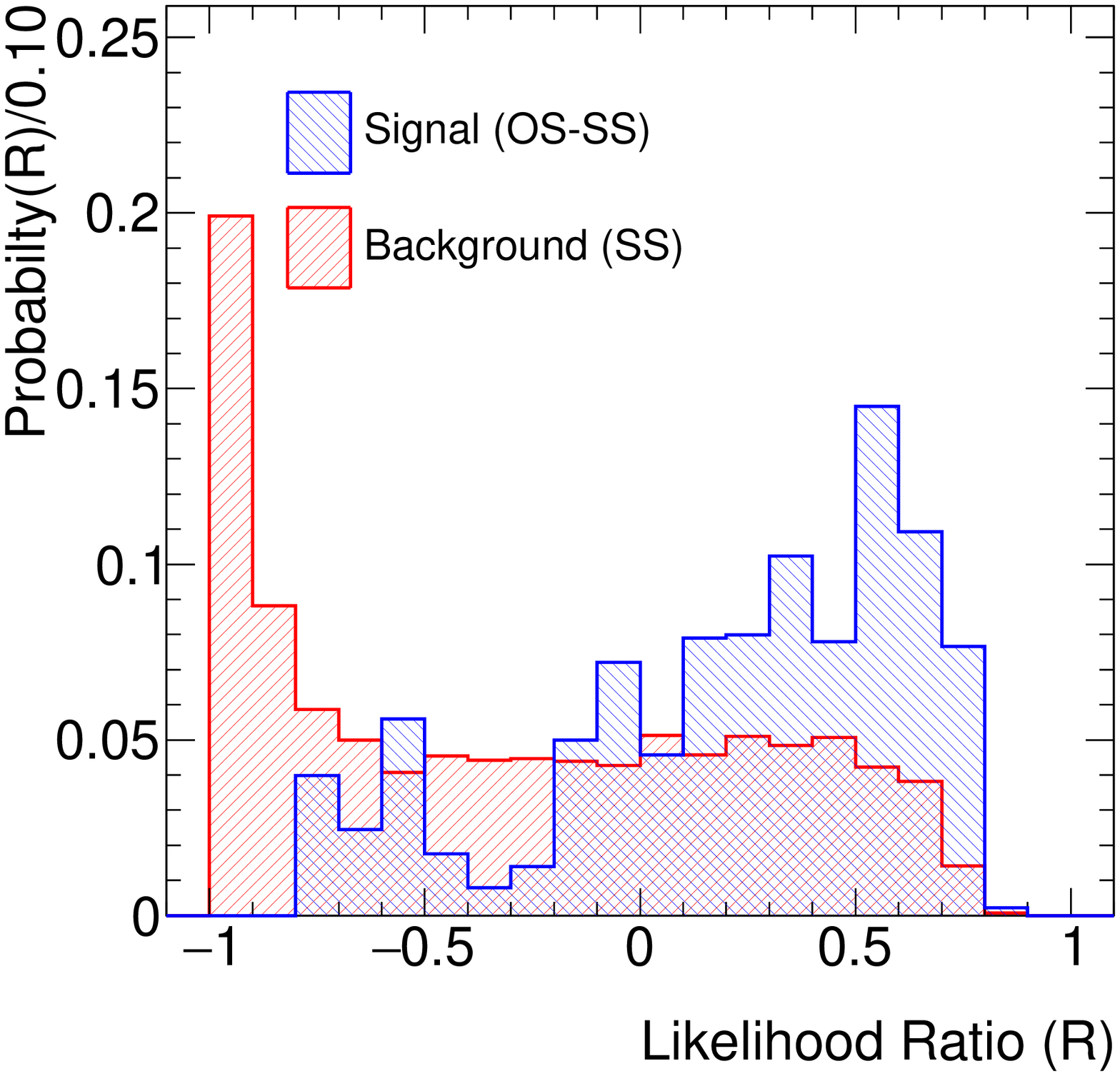}
    } 
    \subfigure[]{
      \label{fig:mu2_best_R_cut}
      \includegraphics[width=0.4\textwidth]{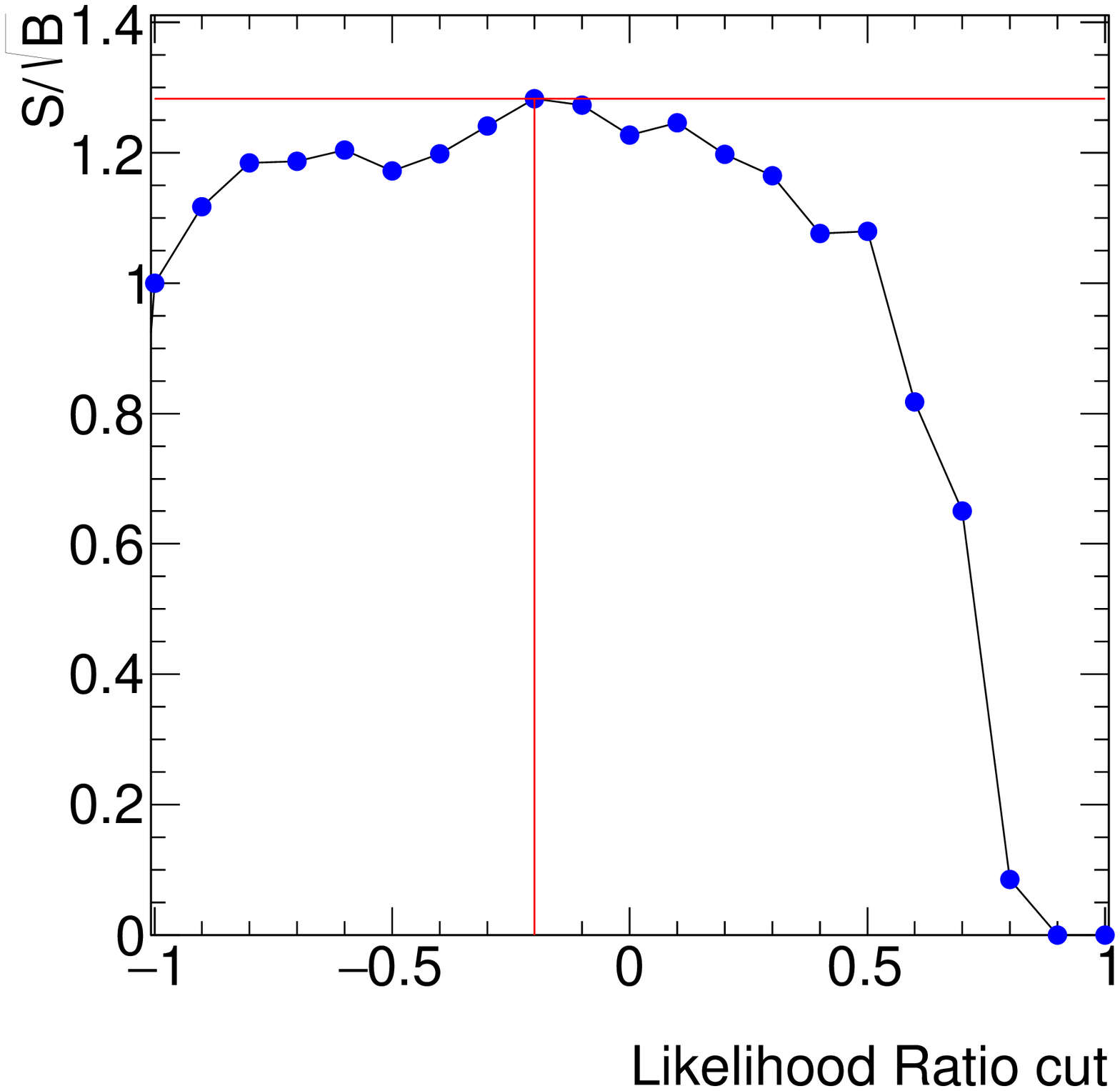}
    }
  \end{center}
  \caption{ \subref{fig:mu2_Likelihood_ratio} Distributions of the Likelihood Ratio (R) for signal (blue) and background (red). \subref{fig:mu2_best_R_cut} The cut value on R variable is optimized using the signal significance.  
  \label{fig:LHratio_muid}}
\end{figure*}

  \end{itemize}

\subsection{Systematic uncertainties}
The following sources of systematic uncertainty on the muon identification efficiency are considered for these three methods: the background and signal fit models used in data, and the smearing on muon $p_T$ in MC. %and matching of $p_T$ resolution of tracks between MC and data. 
%In addition, the background PDFs from sideband regions and using different methods to extract the ratios are also considered in the Likelihood Ratio method. 
For the Likelihood Ratio method, different methods to build the PDF ratios and different procedures to extract the ratios are also considered.
%We used several signal (a single Gaussian or a Crystal-Ball function) and background (a third-order or a fourth-order polynomial function) models to extract number of $J/\psi$, and the fit model uncertainty is determined by using the maximum deviation of any fit result from the average. 
%The fit model uncertainty was described in section 4.1.  
The systematic uncertainties from using different fit models, as described in Section 4.1, are about 5 - 7\% in data. 
%The $p_T$ resolution uncertainty in MC is evaluated by varying the track $p_T$ with $\pm 1\sigma$ of the $p_T$-dependent resolution function, as defined $\frac{p_T^{truth} - p_T^{reco.}}{p_T^{truth}}$. 
The $p_T$ smearing uncertainty in MC is evaluated by varying the mean and width in the smearing function to match the mean and width of reconstructed $J/\psi$ mass within $\pm 1\sigma$.
In addition, for the Likelihood Ratio method, we compared the results from using sideband or same-sign data as the background PDFs, from extracting the ratios in the middle region via bin-to-bin interpolation or via fitting with a third-order polynomial function, and from varying the fit function by $\pm 1 \sigma$ in the side region shown as the blue dot-dashed lines in Fig.~\ref{fig:pdf_ratio}. The maximum deviation from the average of these results is assigned as the uncertainty related to determining the PDF ratios.
%as the PDF ratios related uncertainty. 
%The dominant one comes from the fit model, at about 5 - 7\% in data.  
The total systematic uncertainties in different $p_T$ bins are 0.9 - 8.7\% (0.6 - 2.6\%), 0.9 - 10.5\% (0.2 - 2.1\%) and 0.6 - 15.4\% (0.5 - 3.8\%) for straight cuts, N-1 iteration and Likelihood Ratio method in data (MC), respectively.

%% file: results.tex
\section{Results}
\label{results}

\begin{figure}[!htbp]
  \begin{center}
      \includegraphics[width=0.4\textwidth]{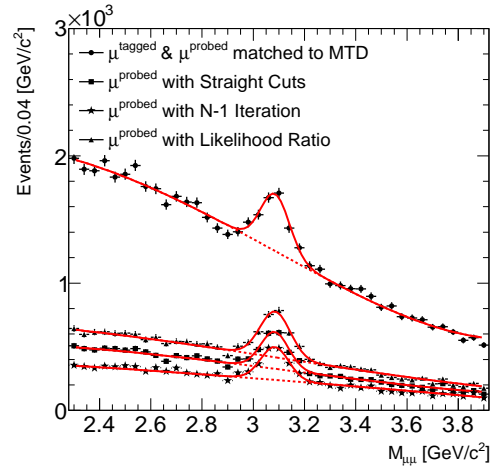}
      \end{center}
      \caption{ The dimuon mass spectra with various muon identification methods. The solid curves are combined fits to the mass distribution with a Gaussian distribution plus a fourth-order polynomial function. The dashed lines show the fitted background. The solid circles are for the basic selection applied to both muons; the open boxes, open circles and solid triangles are for the cases that the subleading muon is selected using straight cut, N-1 iteration cut and the Likelihood Ratio methods, respectively. 
  \label{fig:jpsi_mass_compare}}
\end{figure}

\begin{figure}[!htbp]
  \begin{center}
      \includegraphics[width=0.5\textwidth]{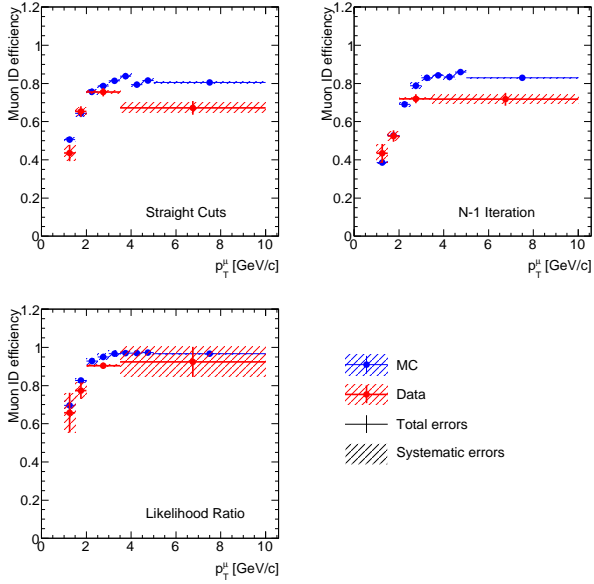}
  \end{center}
  \caption{ The muon identification efficiencies as a function of $p_T$ with straight cuts, N-1 iteration and Likelihood Ratio methods. The blue solid points are from MC, while the red open circles are from data. The error bars are the total uncertainties (statistical plus systematic), while the shaded boxes represent the systematic uncertainties. 
  \label{fig:eff_vs_pt}}
\end{figure}

The performances of different muon identification methods are evaluated by using $J/\psi \to \mu^+ \mu^-$ signals with the selection cuts applied on the probed (subleading) muons as shown in Fig.~\ref{fig:jpsi_mass_compare}. 
%The results demonstrate that all of them have capability to reduce the background level by more than 70\% and to keep the $J/\psi$ signals clean. 
All of them have the capability to reduce the background level by more than 65\% while keeping the $J/\psi$ efficiency relatively high. 
The muon identification efficiencies are calculated relative to the basic selection described in Section 3.2, and shown as a function of $p_T$ in Fig.~\ref{fig:eff_vs_pt}.
The plateau efficiency ($p_T >$ 3 GeV) is about 90\% for the Likelihood Ratio method, and about 80\% for the other methods.  
%Especially the Likelihood Ratio method provides an overall signal efficiency at 74\% and the significance of $J/\psi$ signal is improved by a factor of 1.43. 
For the $J/\psi$ signal, the Likelihood Ratio method provides an overall efficiency of about 80\% and improves the significance by a factor of 1.38.
Detailed comparisons between all three methods are summarized in Table~\ref{table:muid_result}.

\begin{table*}[!htbp]
  \begin{center}
  \begin{tabular}{lcccc}
  \hline\hline
  %\multicolumn{2}{c} { Basic requirements } \\
                              &    Basic selection              &  Straight cut                &  N-1 iteration                & Likelihood Ratio \\ 
  \hline
%  \multicolumn{5}{c} { Data } \\
  Signal                      &     1656 $\pm$ 143 $\pm$ 114    &  1006 $\pm$ 67 $\pm$ 20       &   891 $\pm$ 58 $\pm$ 12       &   1346 $\pm$ 74 $\pm$ 22  \\
  Background                  &     10861 $\pm$ 104 $\pm$ 317   &  2682 $\pm$ 52 $\pm$ 80      &   1962 $\pm$ 44 $\pm$ 58      &    3743 $\pm$ 61 $\pm$ 79  \\
  $S/B$                       &     0.15 $\pm$ 0.01  $\pm$ 0.01 &  0.38 $\pm$ 0.03 $\pm$ 0.01  &   0.45 $\pm$ 0.03 $\pm$ 0.01  &    0.36 $\pm$ 0.02 $\pm$ 0.01 \\
  $S/\sqrt{B}$                &     15.89 $\pm$ 1.37 $\pm$ 1.12 &  19.44 $\pm$ 1.31 $\pm$ 0.49 &   20.11 $\pm$ 1.33 $\pm$ 0.39 &    22.01 $\pm$ 1.23 $\pm$ 0.43  \\
%  $S/\sqrt{S+B}$              &     14.33 $\pm$ 1.06 &  16.31 $\pm$ 1.06   &   16.43 $\pm$ 1.05        &    18.49 $\pm$ 1.06         \\
%  $S/\sqrt{S+2B}$             &     8.87 $\pm$ 0.66  &  11.03 $\pm$ 0.72   &   11.32 $\pm$ 0.72        &    12.69 $\pm$ 0.73         \\
%  $\varepsilon_{signal}$      &     ---              &  0.620 $\pm$ 0.061  &   0.573 $\pm$ 0.056       &    0.739 $\pm$ 0.069         \\
%  $ 1 - \varepsilon_{bkgd.}$  &     ---              &  0.751 $\pm$ 0.014  &   0.799 $\pm$ 0.014        &    0.733 $\pm$ 0.016         \\
  $\varepsilon_{signal}$      &     ---                        &  0.61 $\pm$ 0.01 $\pm$ 0.05   &   0.54 $\pm$ 0.01 $\pm$ 0.05  &    0.81 $\pm$ 0.01 $\pm$ 0.07 \\
  $ 1 - \varepsilon_{bkgd.}$  &     ---                        &  0.75 $\pm$ 0.01 $\pm$ 0.02   &   0.82 $\pm$ 0.01 $\pm$ 0.01  &    0.66 $\pm$ 0.01 $\pm$ 0.01 \\
  \hline
  $\varepsilon_{signal}^{MC}$      &     ---                      &  0.62 $\pm$ 0.01 $\pm$ 0.03   &   0.51 $\pm$ 0.01 $\pm$ 0.03   &    0.80 $\pm$ 0.01 $\pm$ 0.04 \\
%  \multicolumn{5}{c} { Monte Carlo } \\
  \hline \hline
   \end{tabular}
  \end{center}
  \caption{%A summary of the selection criteria for the di-muon candidates.
      Comparison of the performance for three muon identification methods. $\varepsilon_{signal}$ and $ 1 - \varepsilon_{bkgd.}$ are the muon identification efficiency and the background rejection rate relative to the basic selection, respectively. The first and second errors are the statistical and systematic uncertainties, respectively. 
\label{table:muid_result}}
\end{table*}

%\begin{figure}[htbp]
%  \begin{center}
%    \subfigure[]{
%      \label{fig:lr_eff_vs_pt}
%      \includegraphics[width=0.4\textwidth]{efficiency_vs_pt}
%    } 
%    \subfigure[]{
%      \label{fig:lr_sig_vs_pt}
%      \includegraphics[width=0.4\textwidth]{significance_vs_pt}
%    }
%  \end{center}
%  \caption{ \subref{fig:lr_eff_vs_pt} xxxxx.  \subref{fig:lr_sig_vs_pt}
%  \label{fig:lr_result}}
%\end{figure}

\begin{figure}[!htbp]
  \begin{center}
      \includegraphics[width=0.48\textwidth]{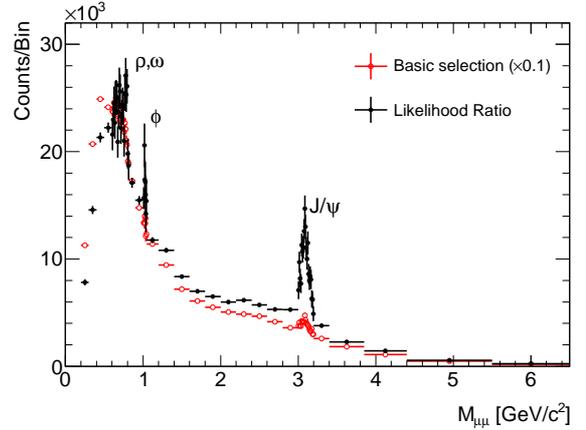}
      \end{center}
      \caption{The dimuon mass spectrum with both muons selected with Likelihood Ratio method compared with that using basic selection which is scaled by 0.1. The significance of $J/\psi$ signal is greatly enhanced and the light mesons are clearly visible. 
  \label{fig:dimuon_mass_after_cut}}
\end{figure}

After applying the muon identification selections determined using the Likelihood Ratio method on both muons, not only the significance of the $J/\psi$ signal is enhanced by a factor of 2 ($S/\sqrt{B} = 31.89$), but also the peaks of the light mesons, such as $\rho$, $\omega$ and $\phi$, become clearer as shown in Fig.~\ref{fig:dimuon_mass_after_cut}. 
This offers a good opportunity to study light mesons, heavy quarkonium, and the dimuon continuum at the STAR experiment.  

%% file: conclusion.tex
\section{Conclusion}
\label{concl}

The MTD is a newly installed detector dedicated to triggering on and identifying muons in STAR with low kinematic cutoff.
%the low energy muons in STAR. 
In this paper, we evaluated three different muon identification methods: straight cut, N-1 iteration and Likelihood Ratio method. Each of these can reduce the background level by more than 65\% and keep the $J/\psi$ signals with about 60\% of efficiency. 
%Each of them has good capability to 
With this muon identification capability, the MTD opens the door to study heavy-ion physics with muons, especially the quarkonium states, in the STAR experiment.

%% file: acknowledgments.tex
\section{Acknowledgments}
\label{ack}

We thank the STAR Collaboration, the RACF at BNL and National Cheng Kung University (Taiwan) for their support. 
This work was supported in part by the U.S. DOE Office of Science under the contract No. de-sc0012704 and the Goldhaber Fellowship program at Brookhaven National Laboratory. 
L. Ruan acknowledges a DOE Office of Science Early Career Award. 
We thank Dr. Gene Van Buren (BNL) for the proofreading.